\documentclass[aps,prl,10pt,twocolumn,longbibliography]{revtex4-2}
\usepackage{graphicx}
\usepackage{amsmath}
\usepackage{amssymb}
\usepackage{mathrsfs}
\usepackage{color}
\usepackage{xcolor}
\usepackage{ulem}
\usepackage{multirow}
\usepackage{epstopdf}
\usepackage[colorlinks=true, citecolor=blue, urlcolor=blue, linkcolor=blue ]{hyperref}

\usepackage[utf8]{inputenc}
\usepackage[T1]{fontenc}
\usepackage[english]{babel}

\makeatletter
\providecommand{\bibnamefont}[1]{#1}
\providecommand{\bibfnamefont}[1]{#1}
\providecommand{\citenamefont}[1]{#1}

\providecommand{\BibitemShut}[1]{}
\makeatother
\begin{document}
	
	\title{Quantum Trajectory Separation and Attosecond Mapping in Liquid High-Harmonic Generation}
	% feichangqi1
	\author{Wanchen Tao$^1$} \email{These authors contributed equally to this work.}
	\author{Ruisi Zhang$^2$} \email{These authors contributed equally to this work.}
    \author{Qihe Guo$^2$}

\author{Lixin He$^{1}$}
\email{helx\_hust@hust.edu.cn}

%%%%%%%%%%%%%%%%%%%%%%%%%%%%%%%%%%%%%%%%
%%%%%%%%%%%%%%%%%%%%%%%%%%%%%%%%%%%%%%%%
\author{Tao-Yuan Du$^{2}$}
\email{duty@cug.edu.cn}

%%%%%%%%%%%%%%%%%%%%%%%%%%%%%%%%%%%%%%%%
%%%%%%%%%%%%%%%%%%%%%%%%%%%%%%%%%%%%%%%%

\author{Xingdong Guan$^1$}

\author{Pengfei Lan$^{1}$}
\email{pengfeilan@hust.edu.cn}

\author{Peixiang Lu$^{1,3}$}
\email{lupeixiang@hust.edu.cn}

\affiliation{%
	$^1$Wuhan National Laboratory for Optoelectronics and School of Physics, Huazhong University of Science and Technology, Wuhan 430074, China\\
	$^2$School of Mathematics and Physics, China University of Geosciences, Wuhan 430074, China\\
	$^3$Hubei Key Laboratory of Optical Information and Pattern Recognition, Wuhan Institute of Technology, Wuhan 430205, China\\
}%

	%\date{\today}% It is always \today, today,
	%  but any date may be explicitly specified
	
\begin{abstract}
High-harmonic generation (HHG) from liquids offers a potential pathway to attosecond spectroscopy in chemically complex and disordered environments, yet fundamental questions remain open: whether liquid harmonic emission preserves well-defined attosecond synchronization, and whether harmonic emission can involve simultaneous contributions from multiple quantum trajectories with distinct excursion times despite strong disorder and scattering. Here, we address these issues experimentally by resolving the trajectory-dependent temporal structure of liquid HHG. By optimizing the laser focusing geometry, we achieve clear spatial discrimination of short- and long-trajectory contributions, providing direct evidence for the existence of multiple quantum trajectories in liquids. Using a phase-controlled two-color driving field, we independently retrieve the attochirp associated with each trajectory and demonstrate opposite energy–time correlations for short and long trajectories, establishing a trajectory-resolved energy–time mapping in liquid HHG. All observations are well reproduced by semiclassical recollision simulations. These results place liquid HHG on the same conceptual footing as gas- and solid-phase HHG and establish a robust foundation for attosecond-resolved spectroscopy of ultrafast electronic and chemical dynamics in liquid environments.

\end{abstract}                         
	
\maketitle

High-harmonic generation (HHG) driven by intense femtosecond laser fields provides unique access to electronic dynamics on attosecond time scales~\cite{corkum1,krausz1}. By coherently upconverting infrared radiation into the extreme-ultraviolet (EUV) regime, HHG underpins attosecond science in two complementary ways. First, it enables the generation of ultrashort attosecond pulses~\cite{Li2017,Gaumnitz2017,25as,19as}, which serve as light sources for time-resolved pump–probe spectroscopy~\cite{Goulielmakis2010,Gallmann2012,Calegari2014,Ramasesha2016,Kaldun2016,Li2024}. Second, HHG functions as an intrinsic attosecond probe through the laser-driven recollision of an electron wave packet with its parent system~\cite{Leone2014,Lpine2014,Lin2010}. In this regime, the emitted high-harmonic spectrum directly encodes ultrafast electronic dynamics, providing attosecond temporal information without the need to explicitly generate isolated attosecond pulses—an approach commonly referred to as high-harmonic spectroscopy (HHS).

%%%%%%%%%%%%%%%%%%%%%%%%%%%%%%%%%%%%%%%

%%%%%%%%%%%%%%%%%%%%%%%%%%%%%%%%%%%%%%%%%%%%%%%%%%%%%%%%%%%%%

A key prerequisite for quantitative HHS is a well-defined correspondence between the emitted photon energy and the electron recollision timing. In gases, the energy–time mapping is well established within the semiclassical recollision picture, in which electrons contributing to different harmonic orders recombine with the parent ion at different times \cite{Corkum1993,Lewenstein1994}. This spread of recollision times results in a spectral group delay dispersion of the harmonic emission, commonly referred to as the attochirp \cite{Doumy2009}. Within this framework, electrons can follow distinct quantum trajectories, most notably the short and long trajectories, which contribute to the same harmonic orders but accumulate different phases and temporal characteristics. Because these trajectories are associated with distinct attochirps, the ability to resolve and control these trajectories has become central to a wide range of spectroscopic applications, including the generation of ultrashort attosecond pulses~\cite{Li2017,Gaumnitz2017,25as,19as}, molecular orbital imaging~\cite{Itatani2004,Vozzi2011} and time-resolved tracking of molecular dynamics~\cite{Baker2006,Smirnova2009,Kraus2015,Lan2017,He2022,He2024}. Related concepts have also been successfully extended to solid-state HHG, where high-harmonic emission is governed by electron–hole dynamics in crystalline band structures~\cite{Ghimire2010,VampaPRB,Ghimire2018,Li2021,Li2023,Li2019}. In this context, distinct interband and intraband pathways encode the effects of ultrafast scattering and dephasing processes, providing access to condensed-matter dynamics on sub-cycle time scales~\cite{Luu2015,Vampa2015,Hohenleutner2015,Luu2018,Li2021,Yue2022,Korolev2024UnveilingTR}.

Liquids constitute an intermediate and fundamentally distinct regime between gases and solids. Strong intermolecular interactions, together with structural disorder and dynamical fluctuations, profoundly modify strong-field electron dynamics, leading to electronic states that are neither fully localized nor fully delocalized \cite{March2002}. Only recently has high-harmonic generation from bulk liquids become experimentally accessible \cite{DiChiara2009,Luu2018NC,Svoboda2021,Ferchaud2022,Mondal2023,MondalOE2023,Alexander2023,Yang_2024,Moore2025,Mondal2025AttosecondresolvedPO,Mondal2025}, revealing several characteristic features, including the apparent independence of the cutoff energy on the driving wavelength \cite{Mondal2023}, a broadened dependence on laser ellipticity \cite{Luu2018NC,Mondal2025}, and the emergence of multiple harmonic plateaus \cite{Mondal2025}. These observations have stimulated the development of collision-based recombination models \cite{Mondal2023,Alexander2023,Mondal2025} as well as time-dependent simulations \cite{neufeld2022ab,Zeng2020,Chen2023,xu2025high}, which highlight the central role of electron–environment interactions in liquid-phase HHG.

From this perspective, liquid HHG offers a potentially powerful route toward attosecond-resolved spectroscopy of liquid-phase dynamics. However, whether this promise can be fully realized hinges on several fundamental unresolved questions. In particular, given the intrinsic structural disorder and strong scattering present in liquids, it remains unclear whether different harmonic orders can preserve well-defined, attosecond-synchronized emission timings, as routinely observed in gases and solids. Equally open is the question of whether liquid HHG supports multiple quantum trajectories with distinct excursion times, which is a prerequisite for establishing a one-to-one energy–time mapping on the attosecond scale. These issues lie at the core of attosecond spectroscopy in liquids and have so far eluded definitive experimental verification. Although theoretical studies suggest that dynamic disorder may strongly suppress long-excursion trajectories and wash out their phase information \cite{Zeng2020}, direct experimental evidence demonstrating the survival of such trajectories and their associated attosecond temporal characteristics is still lacking.

\begin{figure}[t]
	\centering
	\includegraphics[width=8.8cm]{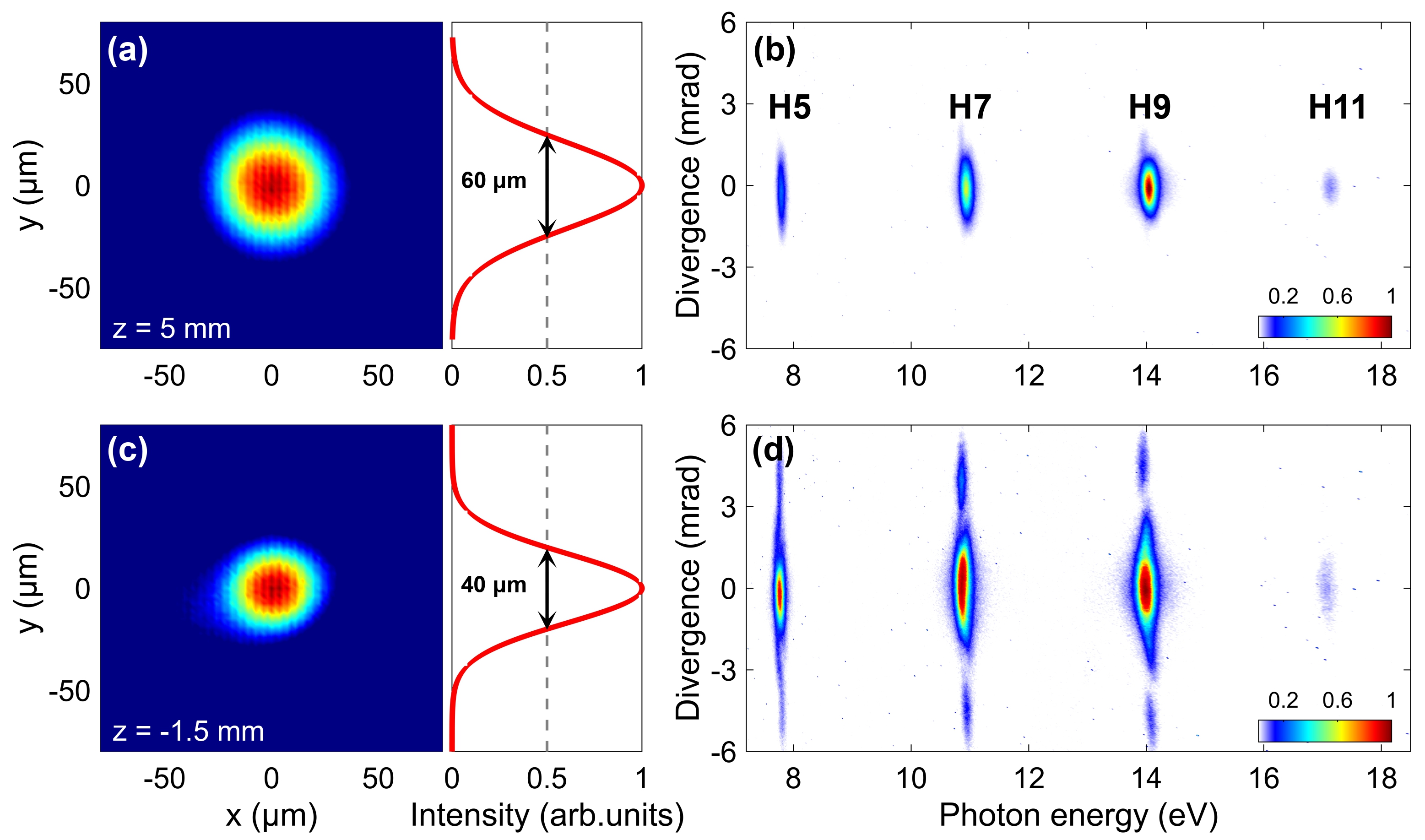}
	\caption{(a) Experimentally measured laser focal spot profile at the position $z = 5$ mm. (b) Corresponding spatial distribution of the generated liquid high-order harmonics. (c)-(d) Same measurements performed at $z =$ $-$1.5 mm. Here, $z>0$ and $z<0$ denote the liquid sheet placed behind and before the focal plane, respectively.}
	\label{fig1}
\end{figure}

Here, we address these fundamental questions experimentally. By optimizing the laser focusing geometry at the generation plane, we resolve distinct spatial signatures associated with short and long quantum trajectories in liquid-phase HHG. To the best of our knowledge, this constitutes the first experimental  evidence that multiple quantum trajectories, including long-excursion trajectories, can survive and contribute to high-harmonic emission in liquids. Leveraging this trajectory-selective control, we further applied a phase-controlled two-color driving field to independently access and quantify the attochirp associated with each trajectory. The resulting trajectory-dependent energy–time mappings establish a comprehensive quantum-trajectory framework for liquid HHG, bridging the conceptual gap between gas-, liquid-, and solid-phase HHS. This advance lays a robust foundation for the extension of attosecond-resolved HHS to chemical reactions and ultrafast dynamics in complex and disordered liquid environments.

%%%%%%%%%%%%%%%%%%%%%%%%%%%%%%%%%%%%%%%%%%%%%%%%%%%%%%%%%%%%%%%%%%%%%%
%%%%%%%%%%%%%%%%%%%%%%%%%%%%%%%%%%%%%%%%%%%%%%%%%%%%%%%%%%%%%%%%%%%%%%
%%%%%%%%%%%%%%%%%%%%%%%%%%%%%%%%%%%%%%%%%%%%%%%%%%%%%%%%%%%%%%%%%%%%%%
%%%%%%%%%%%%%%%%%%%%%%%%%%%%%%%%%%%%%%%%%%%%%%%%%%%%%%%%%%%%%%%%%%%%%%

\begin{figure}[t]
	\centering
	\includegraphics[width=8.5cm]{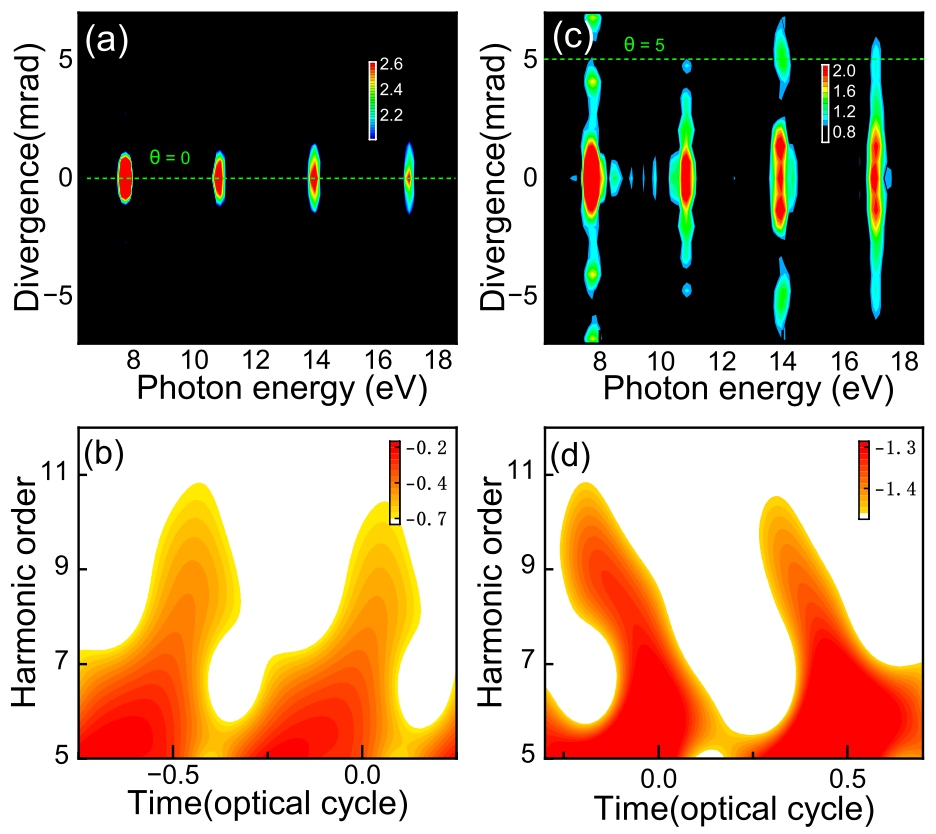}
	\caption{(a) Far-field spatially resolved harmonic spectrum simulated for the large beam size shown in Fig.~\ref{fig1}(a). (b) Time–frequency spectrogram of the on-axis harmonic emission corresponding to panel (a). (c) Same as  (a), but for the small beam size shown in Fig.~\ref{fig1}(c). (d) Time–frequency spectrogram of the off-axis harmonic emission at $\theta = 5$ mrad in (c).}
	\label{fig2}
\end{figure}

In gas-phase HHG, it is well established that the total harmonic phase arises from the combined contributions of the focusing-induced Gouy phase of the driving laser field and the intensity-dependent classical action accumulated along distinct electron trajectories~\cite{Lewenstein1994,Lewenstein1995,Salires1995,Lee2001,Ye2014}. This phase interplay gives rise to characteristic differences in the wavefront curvature of harmonics emitted from short and long trajectories, which manifest as distinct angular divergences in the far field. In particular, near-axis harmonic emission with small divergence is typically associated with short trajectories, whereas more divergent off-axis emission is commonly attributed to long trajectories~\cite{Ye2014,Gaarde1999,Bellini1998,Lyng1999}. Experimentally, such trajectory-dependent spatial separation can be achieved in a simple and robust manner by adjusting the position of the gas target relative to the laser focus.

\begin{figure}[hbt]
	\centering
	\includegraphics[width=8.4cm]{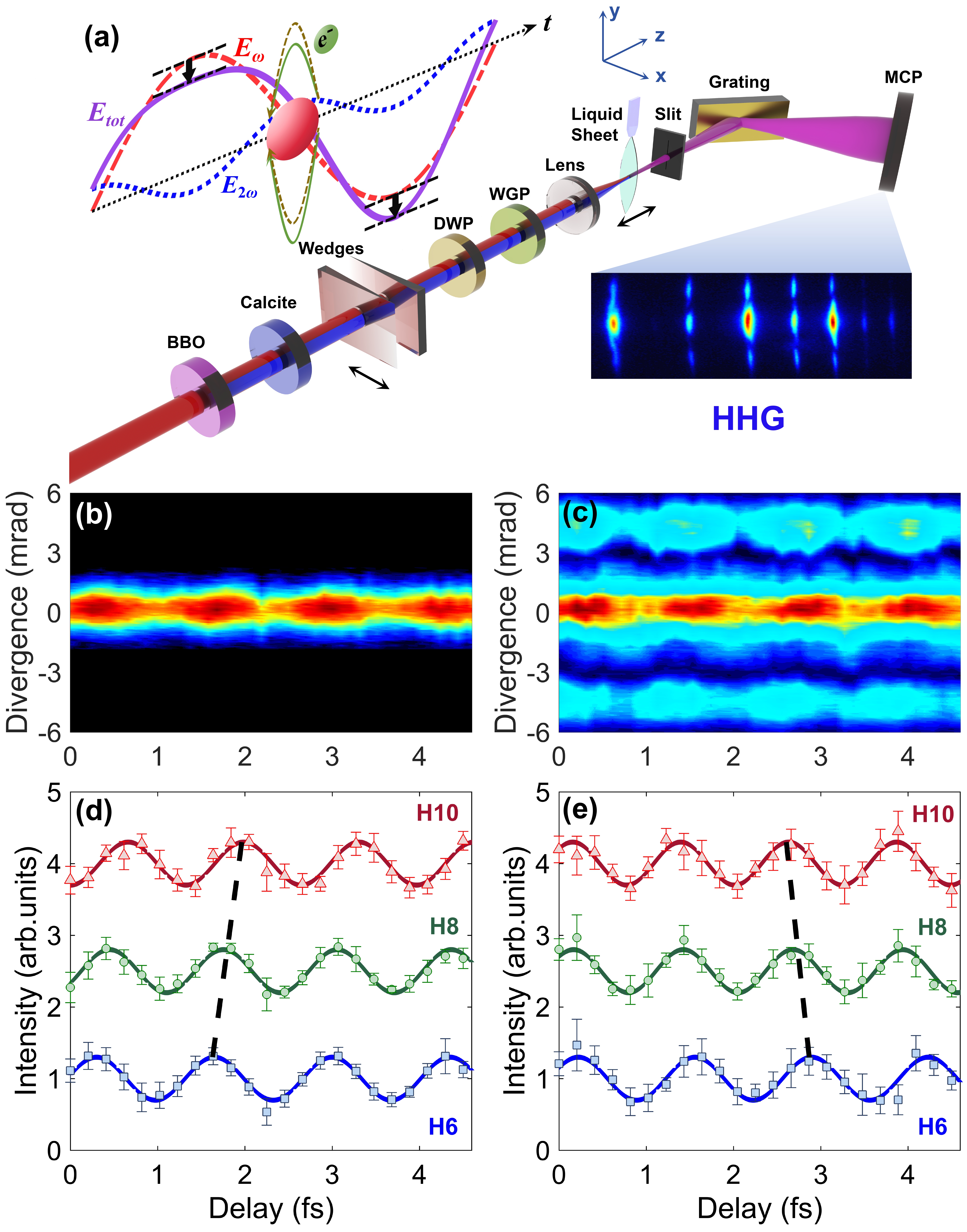}
	\caption{(a) Schematic of the experimental setup of the two-color driving scheme. (b) The spatial distribution of the 8th harmonic (H8) measured at the large beam spot in Fig. \ref{fig1}(a) as a function of the  time delays of the two-color field. (c) Same as (b), but for the results measured at  small beam spot in Fig. \ref{fig1}(c).
    (d) Spatially integrated harmonic intensities of even harmonics (H6-H10) from the short trajectory measured as a function of the two-color time delays. (e) Same as (d), but for the long trajectory. In (d)-(e), the solid lines represent the Fourier fitting of the experimental data.}
	\label{fig3}
\end{figure}

Motivated by the established picture of trajectory-dependent spatial separation in gas-phase HHG, we investigate whether a similar separation of quantum trajectories can be realized in liquid-phase HHG. To this end, we perform HHG measurements from liquids while systematically varying the position of a liquid flat sheet along the laser propagation direction, thereby modifying both the laser intensity distribution and the radial phase profile at the interaction region. In our experiment, 800-nm, 50-fs multi-cycle laser pulses are focused onto a $\sim$1-$\mu$m-thick liquid flat sheet produced by a microfluidic chip (Micro2, Micronit, Inc.). The emitted harmonic radiation is analyzed using a home-built extreme-ultraviolet spectrometer comprising a slit, a flat-field grating, and a microchannel plate detector coupled to a phosphor screen, with the harmonic images recorded by a charge-coupled device camera, as illustrated in Fig.~\ref{fig3}(a).

Figure~\ref{fig1} presents representative spatially resolved harmonic spectra from liquid H$_2$O recorded at different target positions along the laser propagation direction ($z$ axis), where $z=0$ corresponds to the liquid flat-sheet located at the laser focus. When the liquid sheet is positioned well behind the focal plane ($z=5$ mm), corresponding to a relatively large laser spot size (FWHM diameter $D=60~\mu$m), the harmonic emission is predominantly confined to a narrow angular range near the optical axis [Figs.~\ref{fig1}(a)–\ref{fig1}(b)]. As the target is moved closer to the focal region ($z=-1.5$ mm), where the beam spot becomes smaller ($D=40~\mu$m), additional harmonic emission emerges at significantly larger divergence angles, while a low-divergence central component remains visible [Figs.~\ref{fig1}(c)–\ref{fig1}(d)]. Similar divergence patterns are also observed in experiments on liquid ethanol (see Supplementary Materials). The coexistence of low- and high-divergence harmonic components closely resembles the behavior of gas-phase HHG. By analogy with gas-phase HHG studies, the near-axis emissions are expected to arise  from short-trajectory contributions, while the highly divergent off-axis harmonics are more likely associated with long-trajectory emission.

\begin{figure}[hbt]
	\centering
	\includegraphics[width=8.8cm]{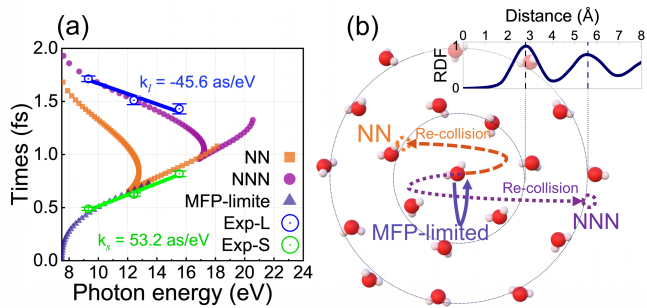}
\caption{(a) Comparison between the experimental and simulated time–frequency mapping of liquid HHG. Open circles with error bars denote the experimental data for the short (green) and long (blue) quantum trajectories. Solid lines represent linear fits to the experimental data, yielding attochirps of $53.2~\mathrm{as/eV}$ and $-45.6~\mathrm{as/eV}$, respectively. Solid symbols indicate the calculated trajectories corresponding to three microscopic mechanisms: mean-free-path-limited (MFP-limited, triangles), nearest-neighbor (NN, squares), and next-nearest-neighbor (NNN, dots) recombination. 
(b) Schematic illustration of the microscopic electron trajectories in the liquid environment. The arrows depict the recollision pathways associated with the MFP-limited (dark blue solid line), NN (orange dashed line), and NNN (purple dotted line) mechanisms shown in (a). The inset displays the oxygen–oxygen radial distribution function of liquid water, with the first and second solvation shells indicated.}
\label{fig:trajectory_analysis}
	\label{fig4}
\end{figure}

To further verify the experimental separation of  the short  and long trajectories, we performed simulations of HHG from liquid water under the same laser conditions as in the experiment. The near-field HHG was obtained from time-dependent Schr\"odinger equation simulations, and its propagation to the far field was computed using a Fresnel diffraction formalism equivalent to a zeroth-order Hankel transform, which captures diffraction-induced beam divergence (see Supplementary Material for details). Figure \ref{fig2}(a) shows the simulated spatially resolved harmonic spectrum for the larger beam spot at $z=5$ mm. In this case, all harmonics exhibit narrow divergence and are tightly confined near the optical axis, in excellent agreement with the experimental results in Fig.~\ref{fig1}(b). The corresponding time–frequency analysis of the on-axis harmonic signal ($\theta=0$) [Fig.~\ref{fig2}(b)] reveals a positive attochirp of the harmonic emission  dominated by the short quantum trajectory.
We also simulated the far-field harmonic spectrum for the smaller beam spot at $z=-1.5$ mm. As shown in Fig.~\ref{fig2}(c), the harmonics display a much broader divergence, with clearly distinguishable on-axis and off-axis components, consistent with the experimental observations in Fig.~\ref{fig1}(d). Moreover, the time–frequency analysis of the off-axis harmonic emission at $\theta=5$ mrad [Fig.~\ref{fig2}(d)] clearly identifies the contribution from the long trajectory. The excellent agreement between simulation and experiment confirms that quantum-trajectory separation is indeed achieved under our experimental conditions. Taken together, these results establish an effective means for resolving and manipulating  short and long quantum trajectories in liquids, enabling trajectory-resolved strong-field studies in condensed-phase systems.

Building on the trajectory-separation strategy established above, we further investigate the trajectory-resolved electron dynamics in liquid HHG using a phase-controlled two-color driving field composed of a strong fundamental and a weak second-harmonic component. In a single-color field, the sub-cycle symmetry of the laser-driven dynamics suppresses even harmonics, resulting in a purely odd-harmonic spectrum (as in Fig.~\ref{fig1}). The addition of a weak second-harmonic field slightly breaks this symmetry, leading to the appearance of even harmonics (as illustrated in Fig.~\ref{fig3}(a)). The intensity of each even harmonic depends on the relative delay between the fundamental and second-harmonic fields and reaches a maximum at a delay that reflects the emission time of that harmonic within the laser cycle. By scanning the two-color delay,  the attochirp, i.e., the dependence of emission time on photon energy, can be extracted from the optimal delays of different even harmonic orders \cite{Dudovich2006,Vampa2015}.

The experimental setup of the two-color scheme is shown schematically in Fig. \ref{fig3}(a). The phase-controlled two-color ($\omega$–2$\omega$) driving field consists of a fundamental 800-nm ($\omega$) pulse and its frequency-doubled 400-nm (2$\omega$) component generated in a $\beta$-barium borate (BBO) crystal. The relative time delay between the two colors is adjusted using a calcite plate, and their relative delay is finely controlled by a pair of fused-silica wedges. A dual wave plate (DWP) (a half-wave plate for the 800-nm fundamental field and a full-wave plate for the 400-nm second-harmonic field) in combination with a wire-grid polarizer (WGP) is employed to tune the intensities of the two fields. The wire-grid polarizer (WGP) ensures parallel polarization of the focused $\omega$ and 2$\omega$ fields. In the interaction region, the peak intensity of the fundamental field is about $5\times10^{13}$~W/cm$^2$, while the intensity of the second-harmonic field is maintained at a perturbative level of approximately $4\times10^{10}$~W/cm$^2$.

We have performed the two-color measurements under the two focusing geometries shown in Fig.~\ref{fig1}. Figures \ref{fig3}(b) and \ref{fig3}(c) display the spatial distribution of the 8th harmonic (H8) as a function of the delay between the $\omega$ and $2\omega$ fields. The narrow-divergence pattern in Fig.~\ref{fig3}(b), obtained at $z=5$ mm, corresponds to on-axis emission dominated by short quantum trajectory. Whereas Fig.~\ref{fig3}(c), measured at $z=-1.5$ mm, reveals off-axis harmonic emission associated with long trajectory. In both cases, the even-harmonic signal exhibits clear periodic modulations as a function of the $\omega$–$2\omega$ delay, with a period equal to one optical cycle of the $2\omega$ field.

Figures \ref{fig3}(d)–\ref{fig3}(e) show the spatially integrated intensities of the measured even harmonics (H6–H10) as a function of the two-color delay for the short- and long-trajectory contributions. For the short trajectories measured at $z=5$ mm, the harmonic yield is integrated over the on-axis region from $-2$ to $2$ mrad. In contrast, for the long trajectory measured at $z=-1.5$ mm, only the off-axis region from $3$ to $6$ mrad is integrated. The optimal two-color delay that maximizes each harmonic order is extracted by Fourier fitting of the delay-dependent modulation (see solid lines in Figs. \ref{fig3}(d)–\ref{fig3}(e)). As indicated by the black dashed lines in Figs.~\ref{fig3}(d)–\ref{fig3}(e), the maxima of the short-trajectory even harmonics systematically shift toward larger delays with increasing photon energy, revealing a positive attochirp of HHG from the short-trajectory. By contrast, the long-trajectory harmonics exhibit an opposite trend. This difference directly reflects the distinct temporal characteristics of these two trajectories, manifested as different slopes in the linear relationship between the optimal delay and the emitted photon energy.

In Fig.~\ref{fig4}(a), the attochirps associated with the short and long trajectories are quantitatively retrieved by linear fitting of the harmonic-order-dependent optimal two-color delays. Note that the peak intensity of the 800-nm fundamental field is kept identical at the two target positions to ensure a meaningful comparison between the extracted attochirps. As shown in Fig.~\ref{fig4}(a), a positive chirp of 53.2 as/eV is obtained for the short-trajectory harmonics, indicating that electrons contributing to higher photon energies spend a longer time in the laser field before recollision. In contrast, a negative chirp of $–$45.6 as/eV is retrieved for the long trajectory, consistent with their fundamentally different excursion dynamics.

To elucidate the microscopic electron dynamics underlying the observed attochirps, we have performed semiclassical trajectory simulations to compare with the experimental measurements. Figure~\ref{fig4}(a) shows the time–frequency mapping, where the experimental data for the short (green circles) and long (blue circles) quantum trajectories are superimposed on the simulated electron trajectories. The calculated trajectories are classified into three different microscopic mechanisms according to their spatial excursion ranges, as schematically illustrated in Fig.~\ref{fig4}(b). Guided by the oxygen–oxygen radial distribution function of liquid water shown in the inset of Fig.~\ref{fig4}(b), we identify: (i) mean-free-path (MFP)-limited trajectories (triangles), for which the electron excursion is confined within the effective scattering mean free path; (ii) nearest-neighbor (NN) recombination (squares), corresponding to electrons scattering from and recombining with molecules in the first solvation shell; and (iii) next-nearest-neighbor (NNN) recombination (dots), involving electron excursions to the second solvation shell~\cite{Mondal2025}. The detailed numerical implementation of these trajectory classes is described in the Supplementary Materials.

The comparison shown in Fig.~\ref{fig4}(a) reveals distinct microscopic origins for the two trajectory classes. The measured short-trajectory timings closely follow the rising edge of the NN and MFP-limited simulations, indicating that short-trajectory emission is predominantly governed by electrons undergoing localized recombination within the first solvation shell. In contrast, the measured long-trajectory timings exhibit a clear correspondence with the longer-excursion branch of the NNN trajectories, pointing to a distinct dynamical origin associated with extended electron propagation prior to recombination. Similar results are also observed in the measurements of HHG from liquid ethanol (see Supplementary Materials). Notably, the observation of long quantum trajectories has not been reported in previous studies of liquid-phase HHG~\cite{Mondal2025,Mondal2025AttosecondresolvedPO}. This absence can be understood as a consequence of the competition between the electron excursion time and the disordered liquid environment. In liquids, structural disorder and high molecular density give rise to frequent scattering events, which rapidly suppress electronic coherence. As a result, trajectories involving long excursion times are generally quenched before recombination can occur. However, in the present experiment, the use of an $800$-nm driving field leads to a substantially shorter optical period compared to the mid-infrared drivers employed in earlier work \cite{Mondal2025}. Consequently, the absolute excursion time of long-trajectory electrons is significantly reduced, enabling them to complete the recollision process and contribute coherently to the harmonic emission before decoherence induced by scattering becomes dominant.

We further note that a recent study reported a positive attochirp of $124 \pm 42$ as/eV for short-trajectory HHG from liquid water driven by a 1800 nm laser~\cite{Mondal2025AttosecondresolvedPO}, approximately 2.3 times larger than the value obtained here. Since the harmonic cutoff energy in liquid HHG is largely insensitive to the driving wavelength, the attochirp is expected to scale approximately linearly with the laser wavelength (or optical cycle) at a fixed intensity. Within this scaling framework, our results are consistent with the earlier report~\cite{Mondal2025AttosecondresolvedPO}. 

%More importantly, the ability to spatially resolve and independently characterize both short and long quantum trajectories in liquids provides a more complete picture of strong-field electron dynamics in liquid-phase HHG.

In summary, we have experimentally demonstrated the trajectory-resolved high-harmonic generation in liquids. By controlling the laser spot size and wavefront curvature at the generation plane, we achieved clear spatial separation of short and long quantum trajectories in liquid-phase HHG, providing the first direct evidence for the survival of long trajectories in a liquid environment. Building on this separation, we retrieve the attochirp associated with each trajectory. The observation of opposite chirp signs for short and long trajectories demonstrates that liquid HHG supports a well-defined trajectory-dependent energy–time mapping, despite the presence of strong intermolecular interactions and dynamic disorder. All the experimental results are well explained by the semi-classical trajectory analysis. These results resolve a long-standing open question in liquid HHG and place liquid-phase HHS on a firm conceptual foundation comparable to that established in gases and solids.

Looking ahead, the trajectory-resolved framework demonstrated here opens several promising directions for liquid-phase attosecond science. In particular, long quantum trajectories, owing to their extended excursion times and propagation distances, are expected to be highly sensitive to electron–environment interactions. Although such trajectories can be strongly suppressed in liquids driven by mid-infrared fields due to wave-packet spreading and frequent scattering, the use of shorter-wavelength driving fields enables long-trajectory recollisions to survive in a disordered liquid environment. This capability provides new access to trajectory-dependent dephasing, electron–hole collision distances, and scattering-induced decoherence, and may offer a route to probing characteristic length scales such as the electron mean free path in liquids. More broadly, combining trajectory separation with wavelength-tunable or waveform-shaped driving fields may enable systematic control of electron excursion time and return energy, providing new handles for disentangling structural and dynamical contributions to liquid-phase HHG spectra. Extending this approach to chemically reactive or heterogeneous liquid systems could ultimately establish attosecond HHS as a general tool for tracking ultrafast electronic and chemical dynamics in complex liquid environments.

%More importantly, the ability to disentangle and control multiple quantum trajectories provides direct access to trajectory-specific electron dynamics in liquids, offering a sensitive probe of continuum propagation, scattering, and decoherence in a condensed yet disordered medium.

%Looking forward, the trajectory-resolved framework demonstrated here opens several promising directions. First, long trajectories—owing to their extended excursion times—are expected to be particularly sensitive to electron–environment interactions, making them a powerful probe of ultrafast solvation dynamics, transient localization, and many-body scattering effects in liquids. Second, combining trajectory separation with wavelength-tunable or waveform-shaped driving fields may enable systematic control of electron excursion time and return energy, providing new handles for disentangling structural and dynamical contributions to liquid HHG spectra. Finally, extending this approach to chemically reactive or heterogeneous liquid systems could establish attosecond HHS as a general tool for tracking ultrafast electronic and chemical dynamics in complex liquid-phase environments.

%%%%%%%%%%%%%%%%%%%%%%%%%%%%
%%%%%%%%%%%%%%%%%%%%%%%%%%%%

\begin{acknowledgements}

This work was supported by the National Key Research and Development Program of China (Grant No. 2023YFA1406800); National Natural Science Foundation of China (Grants No. 12225406, No. 62522505, No. 12474342, and No. 12450406).
\end{acknowledgements}

% \begin{thebibliography}{99}
% 	\bibitem{Allen} L. Allen, M. W. Beijersbergen, R. J. C. Spreeuw, J. P. Woerdman, Orbital angular momentum of light and the transformation of Laguerre-Gaussian laser modes. Phys. Rev. A \textbf{45}, 8185-8189 (1992).
% 	\bibitem{Allen} L. Allen, M. W. Beijersbergen, R. J. C. Spreeuw, J. P. Woerdman, Orbital angular momentum of light and the transformation of Laguerre-Gaussian laser modes. Phys. Rev. A \textbf{45}, 8185-8189 (1992).
	
% 	\bibitem{Korolev} Solids. arXiv:2401.12929 (2024).

% \end{thebibliography}		

%% \bibliography{reference}% 

%apsrev4-2.bst 2019-01-14 (MD) hand-edited version of apsrev4-1.bst
%Control: key (0)
%Control: author (8) initials jnrlst
%Control: editor formatted (1) identically to author
%Control: production of article title (0) allowed
%Control: page (0) single
%Control: year (1) truncated
%Control: production of eprint (0) enabled
%

\end{document}